\documentclass[sn-basic]{sn-jnl}

\usepackage{amssymb}
\usepackage{booktabs}
\usepackage{amsmath, amsthm, tikz, float, subcaption, url}
\usepackage{tabularx}
\usepackage{array}
\usepackage{multirow}
\usepackage[dvipsnames]{xcolor}
\usepackage{graphicx}

\newcolumntype{Y}{>{\raggedright\arraybackslash}X}
\newcommand{\tableformat}{%
    \small
    \renewcommand{\arraystretch}{1.10}
    \setlength{\tabcolsep}{3pt}
}

\begin{document}

\title{Multiperiod bond portfolio optimization with transaction costs using a Markov Decision process}

\author[1]{\fnm{Balaji} \sur{Ramachandran}}
\author[1]{\fnm{Srikanth} \sur{Iyer}}
\author*[2]{\fnm{Shashi} \sur{Jain}}\email{shashijain@iisc.ac.in}

\affil[1]{\orgdiv{Department of Mathematics},
          \orgname{Indian Institute of Science},
          \orgaddress{\city{Bangalore}, \postcode{560012}, \country{India}}}
\affil[2]{\orgdiv{Department of Management Studies},
          \orgname{Indian Institute of Science},
          \orgaddress{\city{Bangalore}, \postcode{560012}, \country{India}}}

\abstract{Bank treasury portfolios must balance yield, liquidity, and interest-rate risk across bonds of different maturities. Static allocation rules are ill-suited to this task: portfolios concentrated in long-duration securities with no dynamic adjustment mechanism can accumulate large mark-to-market losses and liquidity stress under rising interest rates, as illustrated by the failure of Silicon Valley Bank in 2023.

We develop a tractable simulation-based framework for multi-period bond portfolio optimization under interest-rate risk and proportional transaction costs. Yield-curve dynamics are modeled using the Dynamic Nelson-Siegel parameterization with Vector Autoregressive factor dynamics, from which we construct a time-inhomogeneous discrete-state Markov chain approximating the joint yield process across bond maturities. This chain forms the state space of a finite-horizon Markov Decision Process in which the investor maximizes expected terminal wealth subject to proportional rebalancing costs. The optimal portfolio policy is obtained by backward induction. We also quantify the approximation error introduced by truncating the transition kernel, and show that it leaves mean terminal wealth almost unchanged while substantially distorting drawdown and tail statistics.

In simulations using the VAR parameters estimated by \citet{PericoliTaboga2016}, the dynamic programming strategy outperforms both an equal-weight ladder and a static mean-variance portfolio in mean and median terminal wealth. Over six months its advantage over the static mean-variance benchmark ranges from 0.7 percentage points at a transaction cost of 100 basis points to 4.0 percentage points when trading is free, rising to 2.7--9.0 percentage points over twelve months, and it holds across upward-sloping, near-flat and inverted initial yield curves. The gain is a return gain rather than a reduction in risk: because the objective is risk-neutral expected terminal wealth, the dynamic policy does not systematically improve fifth-percentile wealth or maximum drawdown relative to the static benchmarks. As transaction costs rise the strategy trades substantially less, with cumulative turnover falling by roughly five-sixths, but it does not collapse onto the static benchmark and retains an economically material advantage at the highest cost we consider.}

\keywords{Multi-period portfolio optimization, Markov Decision Processes, Dynamic Nelson-Siegel, Markov chain approximation}

\pacs[JEL Classification]{G11, G21, C61, E43}

\maketitle

\section{Introduction}

Banks allocate treasury portfolios across government securities with different maturities as part of asset-liability management and liquidity-risk control \citep{Choudhry2007ALM,Choudhry2011ALM}. Short-term securities support near-term cash needs and regulatory liquidity requirements, whereas longer-maturity bonds can improve portfolio yield over extended horizons \citep{DeVere2025,Englisch2023}. A diversified maturity structure helps align asset cash flows with liabilities and reduces exposure to interest-rate fluctuations and liquidity shortfalls associated with maturity transformation \citep{Yang2021,BrunnermeierGortonKrishnamurthy2013}. These considerations have become especially important under Basel III, whose Liquidity Coverage Ratio requires banks to hold sufficient high-quality liquid assets (HQLA) to withstand a 30-day stress scenario \citep{BCBS2013,Doerr2024}. Empirical studies show that banks subject to LCR requirements have increased their holdings of Treasuries and other HQLA, while maturity diversification helps balance liquidity, reinvestment, and interest-rate risks \citep{Yankov2020,FuhrerMuellerSteiner2016,GiordanaSchumacher2017}.

Static mean-variance optimization remains a standard benchmark for managing such portfolios. Following modern portfolio theory, expected returns and the covariance matrix of bond returns are estimated over a fixed horizon, and portfolio weights are chosen to maximize expected return for a given level of risk \citep{Markowitz1952,Markowitz1959}. Applying this framework across Treasury maturities produces a transparent and computationally tractable trade-off between yield, liquidity, and interest-rate exposure \citep{EltonGruber1997,Ikpe2023,Chandrasekhar2009,Walder2002}. Static mean-variance allocation remains widely used in treasury and ALM applications because it provides an interpretable benchmark for internal risk management and regulatory analysis \citep{BIS2016}.

\subsection{Dynamic Multi Period Portfolio Allocation With Transaction Costs}

Research on multi period portfolio allocation extends the classical Markowitz framework by accounting for the costs and constraints associated with repeated trading. Early stochastic-control models showed that proportional and market-impact costs lead to no-trade regions and state-dependent rebalancing decisions \citep{Constantinides1986,DavisNorman1990,GarleanuPedersen2013}. Dynamic programming methods have since been used to solve finite-horizon portfolio problems with transaction costs and portfolio constraints \citep{SkafBoyd2009,BrownSmith2011,CaiEtAl2020,PalczewskiEtAl2015}. More recent studies apply Wiener-chaos methods and deep reinforcement learning to approximate these dynamic decisions, while there are some other works that show parameter uncertainty can materially affect utility and motivates robust allocation rules \citep{CousinEtAl2024,JiangEtAl2025,CuiEtAl2024,DeMiguelEtAl2015,BertsimasPachamanova2008}.

Although much of this literature considers equity portfolios, related work has begun to address the specific features of fixed-income allocation. Bond strategies must account for yield-curve dynamics, maturity effects, and the trading costs associated with repositioning across the term structure. \citet{ShimaiMakimoto2023} develop a multi-period bond-allocation framework with stochastic interest rates and transaction costs, while \citet{NakayamaTakahashi2007} use dynamic yield-curve factors to determine bond weights over multiple periods. These studies highlight the importance of combining interest-rate forecasts, transaction costs, and the investment horizon when evaluating dynamic bond portfolios \citep{LynchTan2010,MeiDeMiguelNogales2016}.

On the practical side the need for careful portfolio selection of bonds of different maturities by bank treasuries is reflected through the recent case of Silicon Valley Bank and the subsequent contagion to other banks.

\subsection{The Case Of Silicon Valley Bank}

Silicon Valley Bank (SVB) invested a substantial share of its deposits in long-duration U.S. Treasuries and agency mortgage-backed securities. When the Federal Reserve raised interest rates sharply in 2022, the market value of these assets fell, creating large unrealized losses \citep{FedSVBReview2023,OIGSVB2023}. Although held-to-maturity securities would have paid par value if retained until maturity, SVB's concentration in long-duration assets created immediate liquidity pressure. The bank increasingly invested excess deposits in longer-term securities rather than maintaining them in short-duration assets without adequately hedging its interest-rate exposure or holding sufficient liquid assets \citep{Metrick2024,FedSVBReview2023,OIGSVB2023}. When venture-capital funding weakened, SVB's largely uninsured depositors began withdrawing funds rapidly. The bank was forced to sell available-for-sale securities at a realized loss of approximately \$1.8 billion, intensifying depositor concerns \citep{FedSVBReview2023,Metrick2024}.

The episode exposed how concentrated investments in long term securities with interest rate risk can interact to create severe liquidity stress. Signature Bank experienced a similar uninsured-deposit run and was subsequently closed by regulators \citep{FDICSignatureSupervision2023,FDICOIGSignature2023}. The broader concern was reflected in Moody's decision to downgrade its outlook for the U.S. banking sector and place several regional banks under review, citing unrealized securities losses and high uninsured-deposit exposure \citep{MoodySBanks2023}. These events highlight the importance of managing bond portfolios under interest-rate risk.

We use the SVB episode to motivate the interest-rate-risk problem studied here, but we are explicit about what our framework does and does not capture. The model in this paper is an asset-side portfolio-choice problem: the investor holds a fixed set of default-free zero-coupon bonds and rebalances across maturities as the yield curve evolves. It does not model liabilities, deposit outflows, the distinction between held-to-maturity and available-for-sale accounting treatment, or regulatory liquidity constraints, all of which were central to the SVB failure. Our contribution is therefore to the maturity-allocation component of the treasury problem, and the framework should be read as one input to asset-liability management rather than a model of bank funding fragility. Section~5 discusses how liabilities and liquidity constraints could be incorporated.

\subsection{Modelling interest rate fluctuations}

Interest rate dynamics and the evolution of the term structure have been a central focus in fixed-income portfolio management. Early work modeled the short rate directly, with models such as Vasicek and Cox-Ingersoll-Ross providing mean-reverting stochastic dynamics useful for bond pricing, while Hull-White and Black-Karasinski extensions enabled calibration to observed yield curves \citep{Vasicek1977, CoxIngersollRoss1985, HullWhite1990, BlackKarasinski1991, HullWhite1993}.

Term structure modeling subsequently shifted toward multi-factor frameworks, most notably the infinite dimensional Heath-Jarrow-Morton (HJM) framework, which directly models the evolution of the entire forward rate curve under no-arbitrage conditions \citep{HeathJarrowMorton1992}. Complementing these approaches, the Nelson-Siegel parametrization introduced a parsimonious yet flexible representation of yield curves. The Dynamic Nelson-Siegel (DNS) parametrization became a workhorse for yield curve forecasting due to its intuitive level, slope, and curvature interpretation \citep{NelsonSiegel1987, DieboldLi2006, DieboldRudebuschAruoba2006, DieboldRudebusch2013}.

Recognizing that static dynamics may fail to capture structural changes observed in real markets, the literature has extended the DNS model to incorporate regime switching---either via unobserved Markov states or macroeconomic influences---to better model shifts in interest rate behavior \citep{XiangZhu2013, Zhu2015, LevantMa2017}. Building on this trajectory, Pericoli and Taboga develop a Markov-switching Dynamic Nelson-Siegel model that allows factor loadings or volatilities to switch across regimes, improving in-sample fit and forecasting accuracy \citep{PericoliTaboga2016}. We take the estimated factor dynamics of \citet{PericoliTaboga2016} as our calibration, but the yield-curve model we actually solve against is a single-regime specification: the factor process \eqref{auto} has no switching component. Extending the state space to include the regime indicator is straightforward in principle and is left to future work.

Tavanielli and Laurini further enhance this methodology by incorporating regime changes and time-varying parameters with Bayesian estimation to model the Brazilian yield curve, providing a flexible framework capable of capturing structural breaks and regime-dependent dynamics \citep{TavanielliLaurini2023}. Their formulation embeds the Nelson-Siegel factors within a regime-dependent vector autoregressive (VAR) system.

\subsection{The MDP Framework}

The aim of this paper is to develop an optimal selection strategy for a portfolio consisting of a fixed set of bonds that depends on the current state of the yield curve, taking into account the mark-to-market value of the portfolio at each intermediate date.

To model the evolution of interest rates, we develop a Markov chain approximation of the dynamic Nelson-Siegel framework calibrated to \citet{PericoliTaboga2016}, simulating the underlying autoregressive process while focusing only on the yields of bonds held in the bank's portfolio. Because bond maturities decrease over time, the resulting yield process forms a time-inhomogeneous Markov chain driven by the underlying factor dynamics \citep{TavanielliLaurini2023, LevantMa2017, XiangZhu2013, DingEtAl2021}.

Building on this yield curve model, we formulate the portfolio optimization problem as a finite-horizon Markov Decision Process (MDP). In the context of multi-period portfolio optimization, MDPs provide a formal framework for sequential investment decisions under uncertainty. Traditional single-period models are myopic and do not account for intertemporal trade-offs or for mark-to-market losses at intermediate dates. MDPs address these limitations by applying dynamic programming to the sequential allocation problem \citep{XiaYu2025, SkafBoyd2009, BrownSmith2011}.

Recent MDP formulations in portfolio optimization include the work of \citet{XiaYu2025}, who cast finite-horizon multi-period mean-variance optimization as an MDP with an augmented state that tracks accumulated returns, enabling iterative solution via Bellman recursions. These frameworks have been further extended through reinforcement learning and deep RL techniques to handle high-dimensional state and action spaces \citep{CuiEtAl2024, JiangEtAl2025}.

In fixed-income settings, MDPs have been successfully applied to problems involving stochastic interest rates and credit risk. For instance, \citet{PerezHodgeLe2016} transform a wealth allocation problem with defaultable bonds into an MDP, embedding utility maximization and credit-risk considerations into dynamic policies. Broader studies confirm that dynamic policies accounting for stochastic interest rates and transaction costs consistently outperform static single-period allocations \citep{PalczewskiEtAl2015, GarleanuPedersen2013}.

In this paper, we consider the finite-horizon problem of allocating wealth across a portfolio of bonds with different maturities. The objective is to maximize expected terminal wealth \footnote{While we choose to maximize terminal wealth, our framework allows other objectives like the power utility.} over the investment horizon under the time-inhomogeneous Markov chain model of the yield curve. At each time step, rebalancing is permitted subject to linear (proportional) transaction costs, following the treatment in \citep{Pun2022, WangLiu2013, MeiDeMiguelNogales2016}. The resulting optimal strategies are obtained via dynamic programming.

\subsection{Benchmarking and Simulation Strategies}
We carry out simulations to verify the efficacy of our methods. We first calibrate an in-homogenous discrete time Markov chain of interest-rate transition for various maturities in the portfolio from the Nelson-Siegel parametrization. Various realizations of the evolution of the interest-rate vector using the Markov chain are then simulated. On each trajectory several portfolio strategies are implemented and compared under various metrics such as the terminal wealth, maximum drawdown and the turnover of each strategy. We analyse the impact of changing the transaction costs and market conditions such as the volatility parameters of the vector autoregressive process.

The contributions of the paper are as follows:
\begin{enumerate}
    \item We formulate multi-period bond portfolio allocation as a finite-horizon Markov Decision Process with proportional transaction costs, in which the investor maximizes expected terminal wealth by choosing among a discrete set of portfolio allocations at each rebalancing date.
    \item We develop a time-inhomogeneous discrete-state Markov chain approximation of the Dynamic Nelson-Siegel model, deriving state-dependent transition probabilities from simulated VAR paths. This construction is model-agnostic in the sense that the state space and transition probabilities can be calibrated to any yield-curve model that can be simulated forward, parametric or otherwise.
    \item We quantify the approximation error introduced by truncating the transition kernel, which is the step that makes the recursion tractable and is usually left implicit. Retaining only the most probable successors of each state biases the conditional volatility of the approximating chain downward, and the size of the bias depends jointly on the number of successors retained and on the discretization width. In our setting the error leaves mean terminal wealth essentially unchanged while materially distorting drawdown and tail statistics, so a chain that appears adequate on first-moment evidence can be badly miscalibrated for risk measurement.
    \item We solve the resulting Bellman recursion by backward induction, producing an optimal policy at each state and time step that responds explicitly to yield-curve movements rather than imposing a fixed rebalancing rule.
    \item We benchmark the dynamic strategy against equal-weight and static mean-variance portfolios across a range of transaction costs, investment horizons, and yield-factor volatility regimes, quantifying the trade-off between rebalancing frequency and performance. We further show that the result is not an artifact of a particular starting term structure by repeating the comparison under upward-sloping, near-flat and inverted initial yield curves, and we report paired standard errors throughout, exploiting the fact that all strategies are evaluated on common simulated paths.
\end{enumerate}

The closest antecedents in the literature are \citet{ShimaiMakimoto2023}, who study multi-period bond allocation under stochastic interest rates and transaction costs using a linear rebalancing rule, and \citet{NakayamaTakahashi2007}, who use dynamic yield-curve factors to determine bond weights without explicit transaction costs. Both papers work in continuous time and obtain closed-form solutions, but neither models the discrete-time problem with state-dependent trading decisions under practical transaction-cost constraints. The broader multi-period portfolio optimization literature \citep{LynchTan2010, MeiDeMiguelNogales2016} addresses equity or generic risky-asset settings and does not account for the maturity structure and yield-curve dynamics specific to bond portfolios. Our framework addresses these gaps directly.

\subsection{Organization of the paper}

Section 2 describes the multi-period bond portfolio optimization problem and its Markov Decision Process formulation. Section 3 describes the Dynamic Nelson-Siegel framework of parametrizing the yield-curve evolution and the associated Markov Chain approximation underlying the MDP in Section 2. In Section 4, we carry out experiments comparing various static and dynamic allocation strategies. We vary transaction costs, investment horizon, volatility and other parameters and present our results. Finally, we conclude in Section 5 with a discussion of the limitations to our approach and future research directions.

\section{Formulating the multi-period portfolio optimization}
Consider a finite horizon optimization problem of maximizing the gains from a portfolio consisting of $K$ bonds over a discrete time horizon of $T$ periods. The bonds are zero-coupon bonds with maturities $\tau_1 < \tau_2 < \ldots < \tau_K$ at time $t = 0$. Maturities $\tau_k$ are measured in years, while the rebalancing grid is indexed by $i = 0,1,\ldots,T$ with a step length of $\Delta$ years; in the experiments of Section~4, rebalancing is monthly and $\Delta = 1/12$. For simplicity, we assume $T\Delta < \tau_1$, so that no bond matures within the investment horizon. Relaxing this is possible but not immediate: a maturing bond must be removed from the investable set, and its redemption proceeds form part of wealth and must be reinvested, either across the surviving bonds or into a money-market account added to the asset menu. Since the reinvestment rule affects terminal wealth, it is a modelling choice rather than a technicality, and we do not pursue it here. Let $S_{i}$ be the $K$-dimensional random vector of the yields of the bonds at time $t = i$. We denote by $S_{i,k}$ the $k$-th component of $S_{i}$. We allow for $S_{i}$ to take a finite set of values from a state space $\mathcal{S}$. Let $p^{i}(s'|s) = \mathbf{P}(S_{i+1} = s' | S_{i} = s)$ be the transition probability from state $s$ at time $t = i$ to state $s'$ at time $t = i+1$. The transition probabilities are time dependent as the time to maturity of all bonds in the portfolio decrease with time. Deriving the time-dependent transition probabilities for the discrete time chain of approximate bond yields in the portfolio from the homogeneous Markov process describing the evolution of the yield curve is explained in Section 3. At $t = 0$, the investor has $1$ unit of currency to invest in the bonds. Denote by $B_{i,k}$ the price of bond $k$ at time $t = i$ when the yield vector is $S_{i}$. The price of a zero-coupon bond is given by:
\begin{equation}
    B_{i,k} = \exp\big(-(\tau_k - i\Delta)\, S_{i,k}\big).
\end{equation}

Denote by $x_{i,k}$ the fraction of wealth invested in bond $k$ at time $t = i$. We suppose that $x_{i,k}$ can take a finite number of non-negative values such that $\sum_{k=1}^{K} x_{i,k} = 1$. Denote this set of allocation vectors as $\mathcal{A}$, the action space. The investor can rebalance the portfolio at the beginning of each time period. Assume that the cost of rebalancing the portfolio from $x_{i-1}$ to $x_i$ at time $t = i$ is given by
\begin{equation}
\phi(x_{i-1}, x_i, s_{i-1}, s_i) = c\sum_{k=1}^{K} |x_{i,k} - x_{i-1,k}'|,
\end{equation}
where $c \geq 0$ is a constant transaction cost per unit of wealth reallocated; the frictionless case $c=0$ is included as a reference point in Section~4. Note that $x^{'}_{i-1,k}$ is the fraction of wealth in bond $k$ at time $t = i$ just before rebalancing, which is different from $x_{i-1,k}$ due to the change in bond values from time $t = i-1$ to time $t = i$. Specifically,
\begin{equation}
x^{'}_{i-1,k} = \frac{x_{i-1,k} B_{i,k}/B_{i-1,k}}{\sum_{j=1}^{K} x_{i-1,j} B_{i,j}/B_{i-1,j}}.
\end{equation}

The investor's objective is to maximize the expected total gain over the time horizon $T$. Let $g(s_i, x_i, s_{i+1})$ denote the returns of the portfolio, i.e., the change per unit portfolio value, when the state changes from $s_i$ at time $t = i$ to $s_{i+1}$ at time $t = i+1$. The gain is given by:
\begin{equation}
g(s_i, x_i, s_{i+1}) = \sum_{k=1}^{K} x_{i,k} \left(\frac{B_{i+1,k}}{B_{i,k}} - 1\right) = \sum_{k=1}^{K} x_{i,k} \left(\frac{B_{i+1,k}}{B_{i,k}}\right) - 1,
\end{equation}
since $\sum\limits_{k = 1}^K x_{i,k} = 1.$
At $t = 0$ denote the optimal value function or the optimal expected profit accumulated across the time horizon $T$ by $J_0(s_0)$ where $s_0$ is the initial state. At $t = i$, $i = 1,2,\ldots T-1$, the optimal value function per unit portfolio value is denoted by $J_i(s_i, x_{i-1}, s_{i-1})$. At $t = T$, the value function is given by $J_T(s_T, x_{T-1}, s_{T-1}) = 0$ for all $s_T, s_{T-1} \in \mathcal{S}$ and $x_{T-1} \in \mathcal{A}$.

We can express the value function at time $t = i$ in terms of the value function at time $t = i+1$ using the principle of dynamic programming. We now state the dynamic programming equation for any time $1 \leq i \leq T-1$ and explain the rationale for the same.

\begin{equation}
    \label{MDP1}
\begin{aligned}
J_i(s_i, x_{i-1}, s_{i-1}) = &\max_{x_i \in \mathcal{A}} \sum_{s_{i+1} \in \mathcal{S}} p^i(s_{i+1}|s_i) \Big[  (1 + g(s_i, x_i, s_{i+1})) \\
&(1 - \phi(x_{i-1}, x_i, s_{i-1}, s_i)) - 1 \\
&+ \rho J_{i+1}(s_{i+1}, x_i, s_i)(1 + g(s_i, x_i, s_{i+1}))(1 - \phi(x_{i-1}, x_i, s_{i-1}, s_i))\Big].
\end{aligned}
\end{equation}

At time $t = 0$, the value function is given by:
\begin{equation}
    \label{MDP2}
J_0(s_0) = \max_{x_0 \in \mathcal{A}} \sum_{s_1 \in \mathcal{S}} p^0(s_1|s_0) \left[ g(s_0, x_0, s_1) + \rho J_1(s_1, x_0, s_0)(g(s_0, x_0, s_1) + 1) \right].
\end{equation}
Note that no transaction cost appears in \eqref{MDP2}: the initial portfolio is assumed to be established without cost, so that all strategies begin from the same unit wealth and differ only in their subsequent trading. This convention is applied uniformly to the dynamic policy and to both static benchmarks.

Suppose the investor has unit wealth at time $t=i$. After paying transaction costs, the wealth at time $t = i$ is given by $(1 - \phi(x_{i-1}, x_i, s_{i-1}, s_i))$ which at time $t = i+1$ would be $(1 + g(s_i, x_i, s_{i+1}))(1 - \phi(x_{i-1}, x_i, s_{i-1}, s_i))$. Hence the one-step P\&L at time $i$ would be $(1 + g(s_i, x_i, s_{i+1}))(1 - \phi(x_{i-1}, x_i, s_{i-1}, s_i)) - 1$. The second term in the dynamic programming equation \eqref{MDP1} is the expected future gain from time $t = i+1$ to time $t = T$. The expected future gain is given by the value function at time $t = i+1$ multiplied by the wealth at time $t = i+1$, which is $(1 + g(s_i, x_i, s_{i+1}))(1 - \phi(x_{i-1}, x_i, s_{i-1}, s_i))$. We discount the future gain to present value using a sensitivity factor, or the continuation parameter $\rho \in (0, 1]$. Note that the recursion coincides with maximization of expected terminal wealth only when $\rho = 1$; for $\rho < 1$ the investor places less weight on gains accruing later in the horizon, and the objective is a discounted sum of period P\&L rather than terminal wealth. We use $\rho = 1$ throughout, so the stated objective and the solved recursion agree; Table~\ref{tab:rho_sensitivity} reports a sensitivity analysis over $\rho$ and shows that the results are insensitive to this choice.

\subsection{Getting the solution strategy}
We solve MDP \eqref{MDP1} and \eqref{MDP2} using backward induction. We start with $t = T$ and move backwards to $t = 0$. At each time step $i > 0$, we compute the value function $J_i(s_i, x_{i-1}, s_{i-1})$ by numerically solving \eqref{MDP1}. The optimal policy at time $t = i$, $x^*_i(s_i, x_{i-1}, s_{i-1})$ is given by the action that maximizes the value function at time $t = i$ for each state. We iteratively solve for $x_i^*(s_i, x_{i-1}, s_{i-1})$ and appropriately rebalance the portfolio to the current optimal allocation. Finally, we compute the value function at time $t = 0$ using \eqref{MDP2} and obtain the optimal policy at time $t = 0$, $x^*_0(s_0)$.

Because the value function is indexed by the current state, the previous action, and the previous state, the backward recursion requires $O(T\,|\mathcal{S}|^2\,|\mathcal{A}|^2)$ operations in the worst case. In the base case the action set contains $|\mathcal{A}| = 6$ allocations and the reachable state space grows from a single initial state to at most $80$ states per date, giving $341$ states in total over the six-month horizon. Solving the recursion for one transaction-cost level takes approximately $29$ seconds single-threaded in a pure Python implementation; the full experimental grid is embarrassingly parallel across seeds and scenarios.

We now describe how to obtain the transition probabilities $p^i$ using simulations of the Dynamic Nelson-Siegel model.

\section{Markov Chain Approximation Of the Dynamic Nelson Siegel Model}
Consider the Dynamic Nelson-Siegel parametrization of the yield curve \citep{PericoliTaboga2016}. We will use it to infer the transition probabilities of the discrete state Markov Chain as required to solve the MDP \eqref{MDP1},\eqref{MDP2}. The yield for a maturity $\tau$ at time $t$ is given by:
\begin{equation}\label{DNS model}
y_t(\tau) = l_t + s_t \left( \frac{1 - e^{-\lambda \tau}}{\lambda \tau} \right) + \kappa_t \left( \frac{1 - e^{-\lambda \tau}}{\lambda \tau} - e^{-\lambda \tau} \right),
\end{equation}
where $l_t$, $s_t$, and $\kappa_t$ are the level, slope, and curvature factors at time $t$, respectively, and $\lambda$ is a decay parameter called the ``loading factor''. The curvature factor is written $\kappa_t$ rather than $c_t$ to avoid a clash with the transaction-cost parameter $c$ of Section~2.
Let $F_t = [l_t, s_t, \kappa_t]^T$ be the vector of factors at time $t$. The evolution of these factors follows a Vector Auto-regressive (VAR) process:
\begin{equation}\label{auto}
F_t = (I-A) \mu + A F_{t-1} + \eta_t, \quad t = 1, 2, \ldots ,T
\end{equation}
Here $A = $ diag$(a_1, a_2, a_3)$ is a diagonal matrix, $\mu$ is the mean vector constant with time, and $\eta_t \sim N(0, \Sigma_{\eta})$ is a vector of white noise error terms with a constant diagonal covariance matrix. Because both $A$ and $\Sigma_\eta$ are diagonal, \eqref{auto} reduces to three independent AR(1) processes; we retain the vector notation for consistency with the source specification, but no cross-factor dynamics are estimated.

We are interested in obtaining the yields of the $K$ bonds at each timestep. Recall the state $S_t$ at time $t$ is the $K$-dimensional vector of yields of the $K$ bonds in the portfolio at time $t$.

We obtain $y_0(\tau)$ for each maturity $\tau$ in the portfolio. Each yield is then mapped to the nearest value in a discrete set of yields $S_0 \in \mathcal{S}$. This discretization allows us to formulate the problem as an MDP. As the yield curve \eqref{DNS model} evolves, for any $0 \leq t \leq T$, we map the $y_t(\tau)$ values for the various maturities in the portfolio to a discrete vector $S_t \in \mathcal{S}$.

To obtain the transition probabilities $p^t(s'|s)$, we simulate the VAR process \eqref{auto} for a large number of runs. For each run, we map the yields $y_t(\tau)$ to the nearest state $S_t \in \mathcal{S}$. We then count the number of times $n_t(s',s)$ that $S_{t+1}$ takes the value $s'$ given that $S_t$ takes the value $s$. We denote by $N_t(s)$ the total number of simulation runs where $S_t$ takes the value $s$. The transition probability is then estimated as:
\begin{equation}\label{transition_prob}
    p^t(s'|s)= \frac{n_t(s',s)}{N_t(s)}
\end{equation}

Two practical points arise in constructing the chain, the first of which turns out to have a material effect on the results and, to our knowledge, is not quantified elsewhere. The empirical conditional distribution over successor bins must be truncated for the recursion to remain tractable: for each state we retain the $M$ most probable successors subject to a minimum probability threshold, and renormalize so that $\sum_{s'} p^t(s'|s) = 1$. The number of successors retained controls a direct trade-off between tractability and fidelity, because truncation removes mass from the tails of the conditional distribution and therefore understates the volatility of the approximating chain. The severity of this effect depends jointly on $M$ and on the discretization width. At a $20$ basis point grid the one-step conditional distribution of the three yields spreads across roughly $390$ occupied bins, so retaining $M = 10$ successors captures only about one third of the conditional mass and roughly halves the conditional standard deviation of each yield. Widening the grid to $40$ basis points concentrates the same distribution into about $120$ bins, so that $M = 25$ successors capture approximately $92\%$ of the mass. We therefore adopt a $40$ basis point grid with $M = 25$: the additional discretization error from the coarser grid is small relative to the tail-truncation error it removes, and the resulting chain reproduces the conditional volatility of the underlying VAR far more closely. We treat the retained mass as a diagnostic of approximation quality and report it here rather than in the results tables, since it is a property of the chain construction and not of any particular portfolio strategy.

The second point is milder. States are capped at a maximum number per date, retaining the most frequently visited. If a simulated path reaches a yield vector whose nearest discrete state has been pruned, the path is mapped to the closest surviving state under the Euclidean metric on $\mathcal{S}$. In the reported specifications this fallback is never triggered on the evaluation paths, and every state reachable at date $t$ has a non-empty outgoing distribution, so the policy is always well defined along simulated trajectories.

For the estimation of parameters of the autoregressive process we refer to \citet{PericoliTaboga2016}.

\begin{figure}[H]
    \centering
    \includegraphics[width=\textwidth]{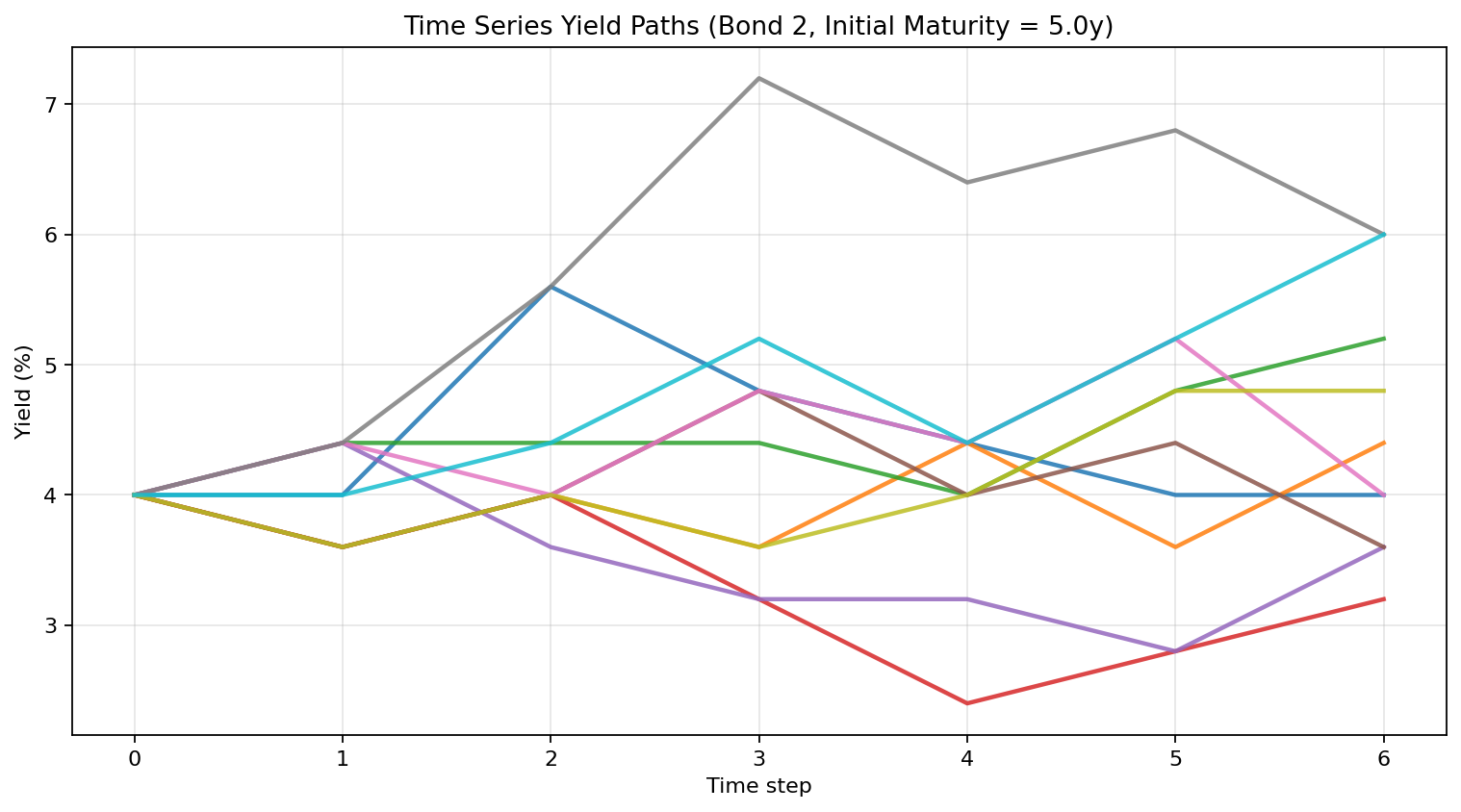}
    \caption{Simulation of Yields of 5 Year Bond over six months using the Markov Chain Approximation with a 40 bps discretization.}
    \label{fig:sim_paths}
\end{figure}

Figure~\ref{fig:sim_paths} illustrates the resulting discretized yield paths for the five-year bond over the six-month horizon.

\section{Experimental Results}
\subsection{Experimental design, benchmark strategies, and evaluation metrics}
We consider the investment of a unit of wealth in a portfolio of three bonds with maturities of 2, 5, and 10 years, respectively. Rebalancing is allowed on a monthly basis over a time horizon of $T = 6$ months.
Yield curve dynamics are modeled using the Dynamic Nelson-Siegel parameterization, with parameters following a Vector Auto-Regressive process as described in Section 3. The parameters of the VAR process are taken from \citet{PericoliTaboga2016}. The state space $\mathcal{S}$ of the Markov Chain is obtained by discretizing the yields $y_t(\tau), 0 \le t \le T$, to the nearest 40 basis points, retaining the $M=25$ most probable successors of each state and capping the number of states at 80 per date, as described in Section~3. We simulate the VAR process \eqref{auto} with the parameters from \citet{PericoliTaboga2016} and map the yields $y_t(\tau)$ obtained from \eqref{DNS model} to the nearest state $S_t \in \mathcal{S}$. The transition probabilities $p^t(s'|s)$ of the Markov Chain are calculated from \eqref{transition_prob}. We consider different levels of transaction costs, $c \in [0, 0.01]$. We also consider three volatility regimes, obtained by scaling the standard deviations of the factor innovations $\eta_t$ in \eqref{auto} about their estimated values.

The action space $\mathcal{A}$ consists of six allocations: the three single-bond corner portfolios, the equal-weight portfolio, and the two intermediate allocations $(0.2,0.3,0.5)$ and $(0.5,0.3,0.2)$. This coarse action set is deliberate, since it keeps the state-action recursion tractable, but it means the dynamic policy is restricted to a small menu of allocations and cannot fine-tune weights continuously. Section~5 discusses relaxing this restriction.

For each parameter configuration, 10,000 sample paths of the Markov Chain are simulated, averaged over five random seeds, except in the initial-curve experiment of Section~\ref{sec:curveshape}, which uses 4,000 paths and three seeds. Every strategy is evaluated on the same simulated paths within each seed, so that the comparisons use common random numbers, reducing Monte Carlo noise across strategies.

Because all strategies are evaluated on common random numbers, we assess the statistical significance of performance differences pathwise. For each simulated path $j$ we compute the difference in terminal wealth $D_j = W_T^{\text{DP}}(j) - W_T^{\text{bench}}(j)$ and report the sample mean $\bar{D}$ together with its standard error $\hat{\sigma}_D/\sqrt{n}$, where $n$ is the number of simulated paths. This paired construction removes the common yield-curve variation and yields far tighter confidence intervals than comparing unpaired means. Paired gaps and their standard errors are reported alongside the level statistics in Table~\ref{tab:curve_shape}.

The base case uses a transaction-cost grid of $c\in\{0,0.0005,0.001,0.002,\\0.005,0.01\}$ and continuation parameter $\rho=1$. The subsequent experiments vary transaction costs, the investment horizon, and the volatility and continuation parameters.

Terminal wealth is the primary performance measure. We report its mean, median, and fifth percentile; the fifth percentile measures severe adverse outcomes and is used as the principal downside-risk statistic. Maximum drawdown is defined, for a simulated path, as the largest percentage decline from a running wealth peak to a subsequent trough. Denoting by $W_t$ the wealth of the portfolio at time $t$, the maximum drawdown is given by:

\begin{equation}
    \text{Max Drawdown} = 100 \times \max_{t \in [0,T]} \left( 1 - \frac{ W_t}{\max_{u \in [0,t]} W_{u}} \right) \%.
\end{equation}

We report the average maximum drawdown across simulated paths. Trading activity is summarized using average turnover and cumulative transaction costs paid. The turnover of a strategy is defined as the total wealth reallocated over the investment horizon:

\begin{equation}
\mathrm{Turnover}
=
\sum_{t=1}^{T}
\sum_{k=1}^{K}
\left|W_{k,t^{+}}-W_{k,t^{-}}\right|.
\end{equation}
 
where $W_{k,t^{+}}$ and $W_{k,t^{-}}$ are the wealth allocated to bond $k$ at time $t$ after and before rebalancing, respectively. 

After presenting the base-case results, we examine how the portfolio policy responds to changes in transaction costs and investment horizon. The transaction-cost analysis examines how trading intensity and strategy performance change as rebalancing becomes more expensive. We also examine whether increasing the horizon to twelve months, which gives each strategy more time to adjust its allocation, preserves the conclusions drawn from the six-month horizon. We also consider alternative volatility regimes and continuation-parameter values to evaluate the robustness of the results to the main modeling assumptions.

We consider two benchmark strategies to compare against the proposed \textbf{Dynamic Programming (DP)} strategy. The benchmark strategies are as follows:
\begin{enumerate}
    \item \textbf{Equal-weight buy-and-hold portfolio}: In this strategy, the investor allocates one-third of wealth to each bond at inception and does not subsequently trade.
    \item \textbf{Static Mean-Variance Optimal Portfolio (MVO)}: In this strategy, the investor allocates the mean-variance optimal composition of wealth to each bond in the portfolio at the beginning of the investment horizon and does not rebalance thereafter. The mean vector and covariance matrix of horizon returns are estimated from an independent set of simulated paths, drawn under the same dynamics but with a different random seed from the paths used for evaluation. The MVO benchmark is therefore not given look-ahead information about the evaluation sample, though it does know the true data-generating process.
\end{enumerate}

In our proposed strategy, \textbf{Dynamic Programming (DP)}, the investor uses the optimal policy obtained from the MDP solution to make decisions at each timestep, over the transaction-cost grid stated above.

\subsection{Main Findings}
The results support four main conclusions.

\begin{enumerate}
    \item First, the DP strategy outperforms both static benchmarks on mean and median terminal wealth in every configuration we examine. Over six months its advantage over equal-weight buy-and-hold ranges from $1.0$ percentage points at a transaction cost of $100$ basis points to $4.3$ percentage points when trading is free, and its advantage over the static mean-variance portfolio from $0.7$ to $4.0$ percentage points. Over twelve months the corresponding ranges are $3.1$--$9.5$ and $2.7$--$9.0$ percentage points. The advantage is therefore largest when trading is cheap and the horizon is long, but it remains economically material at the highest transaction cost considered. Whether the equal-weight ladder loses money depends on the initial curve rather than on the strategy: it does so under the initial factor vector used in earlier drafts, which embeds an expected rise in yields, but not under the curve shapes reported in Section~\ref{sec:curveshape}. The comparison against the static mean-variance benchmark is therefore the more informative of the two throughout.
    \item Second, the gain from dynamic allocation is a return gain, not a risk reduction. Relative to both static benchmarks, DP delivers higher mean and median terminal wealth but does not improve, and frequently worsens, the downside statistics: its fifth-percentile wealth is below that of the static mean-variance portfolio under the upward-sloping, unconditional-mean and original initial curves, and its mean maximum drawdown is of the same order as that of the equal-weight ladder rather than materially smaller. The one exception is the inverted curve, where the mean-variance portfolio concentrates in the ten-year bond and consequently has both the weakest fifth-percentile wealth and the largest drawdown of the three strategies. This is the expected consequence of the objective rather than a defect of the solution: the investor maximizes expected terminal wealth and is risk-neutral by construction, so the policy is never penalized for intermediate mark-to-market losses and will accept additional duration exposure whenever it raises the expected value. A treasury desk facing mark-to-market or capital constraints would require a risk-sensitive objective; the MDP formulation accommodates this directly through a concave utility or a drawdown penalty in the stage reward, which we leave to future work.
    \item Third, higher transaction costs reduce trading intensity but do not extinguish the dynamic policy. Cumulative turnover over six months falls from $6.34$ at $c=0$ to $1.10$ at $c=0.01$, yet the policy still reallocates more than the whole portfolio over the horizon at the highest cost level and retains a $71$ basis point advantage over the static mean-variance benchmark. Cumulative costs paid are not monotone in $c$: they peak at $1.43\%$ of initial wealth at $c=0.005$, where the rising cost rate has not yet been offset by the fall in volume. The dynamic policy therefore moves toward, but does not collapse onto, the static benchmark as trading becomes expensive.
    \item Fourth, the longer horizon increases the opportunity for dynamic allocation, but it also increases potential turnover, so the benefit of DP depends jointly on the investment horizon and transaction costs.
\end{enumerate}

Additionally, we experimented with strategies that rebalance to a fixed static target (equal weighting or the initial MVO weights) at each timestep. These strategies produced significant losses relative to those discussed above and are therefore excluded from the discussion.

\subsection{Comparison across different investment horizons}

Table \ref{tab:horizon_performance} shows the terminal-wealth statistics of each strategy over the six-month investment horizon. The DP strategy has higher mean and median terminal wealth than the other strategies across all transaction costs. However, as transaction costs increase, DP rebalances less frequently, and its advantage over static MVO narrows.

\begin{table}[htbp]
\centering
\tableformat
\caption{Terminal wealth and downside-risk statistics across investment horizons.}
\label{tab:horizon_performance}
\begin{tabular}{llcccc}
\toprule
\textbf{Horizon} & \textbf{Strategy} & \textbf{Mean} & \textbf{Median} & \textbf{Q5} & \textbf{MDD} \\
\midrule
6 months & Equal wt., buy-and-hold & 1.0431 & 1.0408 & 0.9534 & 4.40\% \\
 & MVO, buy-and-hold & 1.0463 & 1.0408 & 0.9616 & 3.93\% \\
 & DP, $c=0$ & 1.0860 & 1.0878 & 0.9500 & 3.92\% \\
 & DP, $c=0.0005$ & 1.0829 & 1.0845 & 0.9477 & 4.01\% \\
 & DP, $c=0.001$ & 1.0800 & 1.0821 & 0.9458 & 4.11\% \\
 & DP, $c=0.002$ & 1.0750 & 1.0776 & 0.9430 & 4.30\% \\
 & DP, $c=0.005$ & 1.0630 & 1.0668 & 0.9317 & 4.42\% \\
 & DP, $c=0.01$ & 1.0534 & 1.0553 & 0.9413 & 3.89\% \\
\midrule
12 months & Equal wt., buy-and-hold & 1.0852 & 1.0820 & 0.9955 & 6.10\% \\
 & MVO, buy-and-hold & 1.0901 & 1.0833 & 1.0032 & 5.87\% \\
 & DP, $c=0$ & 1.1801 & 1.1786 & 1.0196 & 5.65\% \\
 & DP, $c=0.01$ & 1.1166 & 1.1008 & 0.9864 & 5.02\% \\
\bottomrule
\end{tabular}
\begin{flushleft}
\footnotesize
\textit{Notes:} MDD denotes mean maximum drawdown, defined as the largest percentage decline from a running wealth peak to a subsequent trough. Q5 denotes fifth-percentile terminal wealth. All statistics are computed across 10,000 simulated paths using common random numbers and initial wealth normalized to one. The initial yield curve is the unconditional-mean curve of Section~\ref{sec:curveshape}.
\end{flushleft}
\end{table}

From Table \ref{tab:horizon_performance}, we see that increasing the investment horizon to twelve months preserves the ordering of the three strategies on mean and median wealth. The fifth-percentile comparison, however, depends on the horizon: over six months the dynamic policy has lower fifth-percentile wealth than both static benchmarks at every transaction cost, whereas over twelve months it has the highest fifth-percentile wealth of the three at $c=0$. The longer horizon gives the policy more time to trade out of adverse yield-curve states, which improves the left tail; over six months the additional duration exposure it takes on dominates.

Table \ref{tab:horizon_comparison} shows the impact of the extended investment horizon on the performance of the DP and MVO strategies. The twelve-month horizon increases the gap between DP and MVO mean terminal wealth across all values of the transaction-cost parameter. The longer horizon therefore gives DP more time to exploit favorable rebalancing opportunities. However, the increase in trading opportunities also results in greater turnover for the DP strategy over the twelve-month horizon.

\begin{table}[htbp]
\centering
\tableformat
\caption{Comparison of mean wealth and DP turnover across investment horizons.}
\label{tab:horizon_comparison}
\begin{tabular}{lcccc}
\toprule
& \multicolumn{2}{c}{\textbf{Mean wealth}} & \multicolumn{2}{c}{\textbf{DP turnover}} \\
\cmidrule(lr){2-3} \cmidrule(lr){4-5}
\textbf{Strategy} & \textbf{Six months} & \textbf{Twelve months} & \textbf{Six months} & \textbf{Twelve months} \\
\midrule
Equal wt., buy-and-hold & 1.0431 & 1.0852 & -- & -- \\
MVO, buy-and-hold & 1.0463 & 1.0901 & -- & -- \\
DP, $c=0$ & 1.0860 & 1.1801 & 6.339 & 13.822 \\
DP, $c=0.01$ & 1.0534 & 1.1166 & 1.096 & 1.917 \\
\bottomrule
\end{tabular}
\begin{flushleft}
\footnotesize
\textit{Notes:} A dash indicates a static buy-and-hold strategy with zero turnover by construction. Fifth-percentile wealth by strategy and horizon is reported in Table~\ref{tab:horizon_performance}. Turnover is cumulative over the horizon and expressed as a fraction of portfolio wealth, so a value of 6.339 corresponds to reallocating 634\% of wealth in total across the six rebalancing dates.
\end{flushleft}
\end{table}

\subsection{Sensitivity to transaction costs}

Table \ref{tab:trading_activity} shows the turnover and transaction costs of the DP strategy across different values of the transaction-cost parameter. The optimal policy trades less as rebalancing becomes more expensive: cumulative turnover declines monotonically from $6.34$ at $c=0$ to $1.10$ at $c=0.01$, a reduction of roughly five-sixths. The decline is gradual rather than abrupt, and even at the highest cost level the policy continues to rebalance substantially.

Cumulative costs paid, by contrast, are not monotone in $c$. They rise from $0.291\%$ of initial wealth at $c=0.0005$ to a maximum of $1.434\%$ at $c=0.005$, and only then decline, to $1.096\%$ at $c=0.01$. This reflects the two offsetting forces in the product of the cost rate and the volume traded: over most of the grid the rate rises faster than turnover falls, so total costs paid increase; only beyond $c=0.005$ does the reduction in turnover dominate. The implication for practice is that the total cost burden of a dynamic policy is highest at intermediate trading costs, not at the highest ones, where the policy has already curtailed its activity.

\begin{table}[htbp]
\centering
\tableformat
\caption{Trading activity under different transaction costs.}
\label{tab:trading_activity}
\begin{tabular}{lcc}
\toprule
\textbf{Transaction cost $c$} & \textbf{DP turnover} & \textbf{DP cost paid (\% of wealth)} \\
\midrule
0.00\% & 6.339 & 0.000\% \\
0.05\% & 5.815 & 0.291\% \\
0.10\% & 5.310 & 0.531\% \\
0.20\% & 4.566 & 0.913\% \\
0.50\% & 2.869 & 1.434\% \\
1.00\% & 1.096 & 1.096\% \\
\bottomrule
\end{tabular}
\begin{flushleft}
\footnotesize
\textit{Notes:} Transaction costs are expressed in percentage points, so $c=0.10\%$ corresponds to $c=0.001$ in the notation of Section~2. Turnover is cumulative over the six-month horizon. Cost paid is cumulative transaction cost as a percentage of initial wealth.
\end{flushleft}
\end{table}

\subsection{Sensitivity to yield-factor volatility}

Table \ref{tab:volatility_c001} shows the sensitivity of the DP strategy to yield-factor volatility at a transaction cost of $c=0.001$. The results indicate that the DP strategy maintains its advantage over the static strategies in every volatility regime, with higher mean terminal wealth in all three cases. The size of that advantage, however, is remarkably stable: it rises only from $3.32$ to $3.51$ percentage points as the factor standard deviations are scaled from the low to the high regime, and DP mean wealth itself varies by just $44$ basis points across the three regimes. Turnover is likewise close to flat, and not monotone in volatility. What does respond clearly to volatility is downside risk: DP mean maximum drawdown more than doubles, from $2.38\%$ to $5.20\%$, and its fifth-percentile wealth falls from $0.9812$ to $0.9274$. The dynamic policy therefore delivers a similar return advantage across volatility environments while bearing materially more downside risk in volatile ones.

\begin{table}[htbp]
\centering
\tableformat
\caption{Volatility sensitivity at transaction cost $c=0.001$.}
\label{tab:volatility_c001}
\begin{tabular}{llccccc}
\toprule
\textbf{Vol.} & \textbf{Strategy} & \textbf{Mean} & \textbf{Q5} & \textbf{MDD} & \textbf{Turn.} & \textbf{TC paid} \\
\midrule
\multirow{3}{*}{Low} & DP & 1.0794 & 0.9812 & 2.38\% & 5.5824 & 0.005582 \\
 & MVO, buy-and-hold & 1.0462 & 0.9861 & 2.44\% & -- & -- \\
 & Equal wt., buy-and-hold & 1.0422 & 0.9786 & 2.79\% & -- & -- \\
\midrule
\multirow{3}{*}{Baseline} & DP & 1.0800 & 0.9458 & 4.11\% & 5.3096 & 0.005310 \\
 & MVO, buy-and-hold & 1.0463 & 0.9616 & 3.93\% & -- & -- \\
 & Equal wt., buy-and-hold & 1.0431 & 0.9534 & 4.40\% & -- & -- \\
\midrule
\multirow{3}{*}{High} & DP & 1.0838 & 0.9274 & 5.20\% & 5.4410 & 0.005441 \\
 & MVO, buy-and-hold & 1.0487 & 0.9503 & 4.95\% & -- & -- \\
 & Equal wt., buy-and-hold & 1.0456 & 0.9438 & 5.41\% & -- & -- \\
\bottomrule
\end{tabular}
\begin{flushleft}
\footnotesize
\textit{Notes:} Q5 denotes the fifth percentile of terminal wealth. MDD denotes mean maximum drawdown. Turnover and TC paid are reported only for DP; buy-and-hold strategies have zero turnover and zero transaction costs by construction. TC paid denotes cumulative transaction costs as a fraction of initial wealth. Low, baseline, and high volatility correspond to factor standard deviations $(0.20,0.40,0.85)$, $(0.36,0.62,0.92)$, and $(0.50,0.80,0.99)$, respectively, for the level, slope, and curvature factors of \eqref{auto}. The baseline values are those estimated by \citet{PericoliTaboga2016}; the low and high regimes are chosen to bracket them.
\end{flushleft}
\end{table}

We note that the fifth-percentile comparison reverses across regimes: in the low-volatility regime DP has a lower mean maximum drawdown than the static mean-variance portfolio ($2.38\%$ against $2.44\%$), while in the baseline and high regimes it has a higher one. The dynamic policy's risk profile is thus close to that of the static benchmarks when the curve is calm and deteriorates relative to them as volatility rises.

\subsection{Sensitivity to the initial yield curve}\label{sec:curveshape}

The results reported above condition on a single initial factor vector. Because the DNS factors mean-revert, the initial curve determines both the shape of the term structure the investor faces and the direction in which it is expected to move, so it is a first-order determinant of any maturity-allocation result. We therefore re-estimate the model under three initial curves that differ in shape rather than in level. In each case the level and curvature factors are set to their estimated long-run means and only the slope factor is varied, so that no deterministic level drift is imposed: an upward-sloping curve (slope factor $-3.00$), the unconditional mean curve (slope factor $-1.50$, mildly upward-sloping), and an inverted curve (slope factor $+1.50$). For comparability with earlier drafts we also report the original initial factor vector $(5.00, -2.00, 3.00)$, which starts the level factor well below its long-run mean and therefore embeds an expected increase in yields of roughly $115$ basis points over the six-month horizon.

Table~\ref{tab:curve_shape} reports the results. Three points stand out. First, the static mean-variance benchmark is a corner solution in every scenario, allocating all wealth to a single maturity: the two-year bond when yields are expected to rise, the five-year bond at the unconditional mean, and the ten-year bond when the curve is inverted and expected to re-steepen. This is inherent to a long-only problem with three assets whose risk and expected return are both ordered by duration, and it means that ``static mean-variance'' should be read as ``the single best maturity given the initial curve''. Second, the advantage of the dynamic policy over this benchmark is substantially larger than earlier estimates suggested, ranging from $2.0$ to $3.3$ percentage points at a transaction cost of $10$ basis points, and is statistically unambiguous under the paired construction. Third, the extent to which high transaction costs push the dynamic policy toward the static benchmark itself depends on the initial curve. At a transaction cost of $100$ basis points the policy all but stops trading under the inverted and original curves, where cumulative turnover falls to $0.15$ and $0.19$ and the advantage over the static benchmark falls to $0.05$ and $0.10$ percentage points. Under the upward-sloping and near-flat curves it continues to reallocate more than the whole portfolio over the horizon, with turnover of $1.53$ and $1.17$, and retains an advantage of $0.52$ and $0.58$ percentage points. Where a single maturity dominates, the policy holds it; where the curve offers relative-value movement between maturities, trading remains worthwhile even at high cost.

\begin{table}[htbp]
\centering
\tableformat
\caption{Mean terminal wealth by initial yield-curve shape.}
\label{tab:curve_shape}
\begin{tabular}{llccccrc}
\toprule
\textbf{Initial curve} & \textbf{$c$} & \textbf{Equal wt.} & \textbf{MVO} & \textbf{DP} &
\textbf{DP$-$MVO (pp)} & \textbf{$t$} & \textbf{DP turnover} \\
\midrule
 \multirow{3}{*}{Upward sloping} & 0 & 1.0202 & 1.0243 & 1.0612 & +3.69 & 56 & 6.26 \\
  & 0.001 & 1.0202 & 1.0243 & 1.0553 & +3.10 & 47 & 5.23 \\
  & 0.01 & 1.0202 & 1.0243 & 1.0294 & +0.52 & 8 & 1.53 \\
\midrule
 \multirow{3}{*}{At long-run mean} & 0 & 1.0449 & 1.0479 & 1.0878 & +4.00 & 86 & 6.51 \\
  & 0.001 & 1.0449 & 1.0479 & 1.0813 & +3.34 & 75 & 5.56 \\
  & 0.01 & 1.0449 & 1.0479 & 1.0536 & +0.58 & 18 & 1.17 \\
\midrule
 \multirow{3}{*}{Inverted} & 0 & 1.1069 & 1.1430 & 1.1682 & +2.52 & 53 & 5.14 \\
  & 0.001 & 1.1069 & 1.1430 & 1.1629 & +1.99 & 43 & 4.52 \\
  & 0.01 & 1.1069 & 1.1430 & 1.1435 & +0.05 & 5 & 0.15 \\
\midrule
 \multirow{3}{*}{Original paper} & 0 & 0.9825 & 0.9952 & 1.0226 & +2.74 & 52 & 6.27 \\
  & 0.001 & 0.9825 & 0.9952 & 1.0167 & +2.15 & 41 & 5.37 \\
  & 0.01 & 0.9825 & 0.9952 & 0.9962 & +0.10 & 6 & 0.19 \\
\bottomrule
\end{tabular}
\begin{flushleft}
\footnotesize
\textit{Notes:} Level and curvature factors are held at their long-run means in the first three scenarios; only the slope factor varies, so the scenarios differ in curve shape rather than in an imposed level drift. The final scenario reproduces the initial factor vector used in earlier drafts. DP$-$MVO is the mean pathwise difference in terminal wealth in percentage points, and $t$ is the corresponding paired $t$-statistic; the paired construction exploits common random numbers across strategies. Results average three seeds with 4{,}000 paths each, using a $40$ basis point grid with $25$ retained successors and at most $80$ states per date. Tables~\ref{tab:horizon_performance}--\ref{tab:rho_sensitivity} use five seeds with 10{,}000 paths each on the same chain specification, so figures for the unconditional-mean curve differ between the two sets of tables by a few basis points of Monte Carlo error.
\end{flushleft}
\end{table}

Table~\ref{tab:curve_downside} reports the corresponding downside statistics. The dynamic policy does not dominate the static benchmarks on either measure. Its fifth-percentile wealth is below that of the mean-variance portfolio in three of the four scenarios, and its mean maximum drawdown exceeds that of the equal-weight ladder in two of the four. The exception is the inverted curve, where the mean-variance portfolio concentrates in the ten-year bond and records both the weakest fifth-percentile wealth and the largest drawdown of the three strategies. As discussed above, this reflects the risk-neutral objective rather than a failure of the policy.

\begin{table}[htbp]
\centering
\tableformat
\caption{Downside statistics by initial yield-curve shape, transaction cost $c=0.001$.}
\label{tab:curve_downside}
\begin{tabular}{lcccccc}
\toprule
& \multicolumn{3}{c}{\textbf{Fifth-percentile wealth}} & \multicolumn{3}{c}{\textbf{Mean maximum drawdown}} \\
\cmidrule(lr){2-4} \cmidrule(lr){5-7}
\textbf{Initial curve} & \textbf{Equal wt.} & \textbf{MVO} & \textbf{DP} & \textbf{Equal wt.} & \textbf{MVO} & \textbf{DP} \\
\midrule
 Upward sloping & 0.9292 & 0.9960 & 0.9185 & 5.26\% & 1.36\% & 5.16\% \\
 At long-run mean & 0.9511 & 0.9627 & 0.9429 & 4.34\% & 3.88\% & 4.36\% \\
 Inverted & 1.0051 & 0.9622 & 0.9930 & 2.84\% & 6.06\% & 3.67\% \\
 Original paper & 0.8933 & 0.9665 & 0.8974 & 6.72\% & 2.37\% & 5.04\% \\
\bottomrule
\end{tabular}
\end{table}

\subsection{Sensitivity to the continuation-weight parameter}

\begin{table}[htbp]
\centering
\tableformat
\caption{Sensitivity of DP performance to the continuation parameter $\rho$.}
\label{tab:rho_sensitivity}
\begin{tabular}{lccccc}
\toprule
\textbf{$\rho$} & \textbf{Mean} & \textbf{Q5} & \textbf{MDD} & \textbf{Turn.} & \textbf{TC paid} \\
\midrule
\multicolumn{6}{l}{$c=0$} \\
0.90 & 1.0860 & 0.9500 & 3.93\% & 6.34 & 0.000000 \\
0.95 & 1.0860 & 0.9500 & 3.93\% & 6.34 & 0.000000 \\
1.00 & 1.0860 & 0.9500 & 3.92\% & 6.34 & 0.000000 \\
\midrule
\multicolumn{6}{l}{$c=0.01$} \\
0.90 & 1.0533 & 0.9394 & 4.25\% & 1.05 & 0.010496 \\
0.95 & 1.0534 & 0.9416 & 3.84\% & 1.05 & 0.010546 \\
1.00 & 1.0534 & 0.9413 & 3.89\% & 1.10 & 0.010959 \\
\bottomrule
\end{tabular}
\begin{flushleft}
\footnotesize
\textit{Notes:} Q5 denotes the fifth percentile of terminal wealth. MDD denotes mean maximum drawdown. TC paid denotes cumulative transaction costs as a fraction of initial wealth.
\end{flushleft}
\end{table}

As shown in Table~\ref{tab:rho_sensitivity}, changing $\rho$ from $0.90$ to $1.00$ produces no economically meaningful change in DP performance. Mean terminal wealth is identical to four decimal places at $c=0$ and differs by at most one unit in the fourth decimal at $c=0.01$.

This near-invariance is not because the parameter is inert. Varying $\rho$ from $1.00$ to $0.90$ changes the selected action in $0.4\%$ of state-action-history combinations at $c=0$ and in $2.0\%$ at $c=0.01$. Performance is nevertheless almost unchanged because continuation values are small relative to the one-period differences across actions, so the states in which the argmax flips are visited rarely along simulated paths. At $c=0$ the affected states are visited so seldom that turnover is unchanged to two decimal places; at $c=0.01$, where a larger share of the policy is affected, turnover moves from $1.05$ to $1.10$ and mean maximum drawdown from $4.25\%$ to $3.89\%$. These differences are within the range of seed-to-seed variation and we do not interpret them. The results are driven by transaction costs rather than by the continuation parameter, and we use $\rho=1$ as the baseline specification throughout.

\section{Conclusion and Future Work}

Bank treasury portfolios must balance yield, liquidity, and interest-rate risk across bonds of different maturities. Static allocation rules are ill-suited to this task: as the SVB episode illustrated, a portfolio concentrated in long-duration fixed-income securities with no dynamic adjustment mechanism can accumulate large mark-to-market losses and liquidity stress under rising interest rates. This paper develops a tractable dynamic framework for multi-period bond portfolio allocation that responds explicitly to yield-curve movements.

We make three methodological contributions. First, we construct a time-inhomogeneous discrete-state Markov chain that approximates the Dynamic Nelson-Siegel yield-curve model of \citet{PericoliTaboga2016}, using simulated VAR paths to estimate state-dependent transition probabilities. Second, we formulate the bond allocation problem as a finite-horizon Markov Decision Process with proportional transaction costs, in which the investor maximizes expected terminal wealth by choosing among a discrete set of portfolio allocations at each rebalancing date. Third, we solve the resulting Bellman recursion by backward induction, yielding a numerically tabulated optimal policy indexed by each state and time step.

The experimental results support four conclusions. First, the DP strategy outperforms equal-weight buy-and-hold by $1.0$--$4.3$ percentage points in mean terminal wealth over six months and by $3.1$--$9.5$ percentage points over twelve months, and the static mean-variance portfolio by $0.7$--$4.0$ and $2.7$--$9.0$ percentage points respectively, with the range spanned by transaction costs from $100$ basis points down to zero. Second, the advantage is a return advantage rather than a reduction in risk: DP does not systematically improve fifth-percentile wealth or maximum drawdown relative to either static benchmark, which follows directly from the risk-neutral objective. Third, higher transaction costs reduce trading intensity substantially, cumulative turnover falling by roughly five-sixths between zero and $100$ basis points, but the policy does not collapse onto the static benchmark and retains a $71$ basis point advantage at the highest cost level; cumulative costs paid peak at intermediate cost levels rather than at the highest. Fourth, the ordering of the three strategies is unchanged under upward-sloping, near-flat and inverted initial yield curves, so the result does not depend on a particular starting term structure.

The framework has several limitations that point toward future work. The discrete action and state spaces introduce approximation error, and the finite set of portfolio weights may preclude some beneficial adjustments available in continuous-action settings. The transition probabilities are estimated from simulated paths of a specific VAR parameterization \citep{PericoliTaboga2016}, so performance depends on the accuracy of those parameter estimates and the adequacy of the DNS model for the yield curve in question. In particular, all results reported here are obtained under the data-generating process assumed by the model itself, so they measure the value of dynamic rebalancing when the yield-curve dynamics are known, and should be interpreted as an upper bound on what is achievable out of sample. The analysis also assumes zero-coupon bonds and a fixed portfolio composition, and models transaction costs as a simple proportional charge rather than incorporating bid-ask spreads, market impact, or regulatory liquidity requirements. Finally, the objective is risk-neutral expected terminal wealth; a treasury desk subject to mark-to-market or capital constraints would require a risk-sensitive objective, which the MDP framework accommodates but which we do not explore here.

In future work, we plan to address these limitations in three directions. First, and most importantly, we intend to estimate the DNS-VAR system directly on observed Treasury yields and to evaluate the policy out of sample over the 2021--2023 tightening cycle, which would test whether the framework delivers value under realized rather than simulated dynamics. On the modeling side, we aim to incorporate liquidity constraints and more realistic transaction-cost structures, and to extend the framework to coupon-bearing bonds and portfolios whose composition can be adjusted over time, together with a liability side comprising a stochastic deposit-outflow process and an LCR-type constraint, which would bring the model closer to the asset-liability problem the SVB episode exposed. On the computational side, deep reinforcement learning methods offer a natural path to continuous action spaces and higher-dimensional state representations, removing the need for explicit state-space discretization and enabling the framework to scale to larger portfolios of bonds across the full term structure.

\section*{Declaration of competing interest}
The authors declare that they have no known competing financial interests or personal relationships that could have appeared to influence the work reported in this paper.

\section*{Funding}
This research did not receive any specific grant from funding agencies in the public, commercial, or not-for-profit sectors.

\section*{Data and code availability}
No proprietary data were used. All results are generated from simulations of the Dynamic Nelson-Siegel model with parameters taken from \citet{PericoliTaboga2016}. 

\bibliography{references}

@article{Markowitz1952,
  author  = {Markowitz, Harry},
  title   = {Portfolio Selection},
  journal = {The Journal of Finance},
  volume  = {7},
  number  = {1},
  pages   = {77--91},
  year    = {1952},
  doi     = {10.1111/j.1540-6261.1952.tb01525.x}
}

@book{Markowitz1959,
  author    = {Markowitz, Harry M.},
  title     = {Portfolio Selection: Efficient Diversification of Investments},
  publisher = {John Wiley \& Sons},
  address   = {New York},
  year      = {1959}
}

@article{EltonGruber1997,
  author  = {Elton, Edwin J. and Gruber, Martin J.},
  title   = {Modern Portfolio Theory, 1950 to Date},
  journal = {Journal of Banking \& Finance},
  volume  = {21},
  number  = {11-12},
  pages   = {1743--1759},
  year    = {1997},
  doi     = {10.1016/S0378-4266(97)00048-4}
}

@article{Ikpe2023,
  author  = {Ikpe, Dennis and Mawonike, Romeo and Viens, Frederi},
  title   = {Static Markowitz Mean-Variance Portfolio Selection Model with Long-Term Bonds},
  journal = {Numerical Algebra, Control and Optimization},
  year    = {2023},
  doi     = {10.3934/naco.2022030},
  note    = {First published online 2022}
}

@mastersthesis{Chandrasekhar2009,
  author       = {Chandrasekhar, Rohan},
  title        = {Fixed-Income Portfolio Optimization},
  school       = {University of Texas at Austin},
  year         = {2009},
  note         = {Master's Report, Department of Mechanical Engineering}
}

@techreport{Walder2002,
  author      = {Walder, Roger},
  title       = {Dynamic Allocation of Treasury and Corporate Bond Portfolios},
  institution = {International Center for Financial Asset Management and Engineering (FAME)},
  type        = {FAME Research Paper Series},
  number      = {64},
  year        = {2002}
}

@article{Englisch2023,
  author  = {Englisch, Holger and Krabichler, Thomas and M{\"u}ller, Konrad J. and Schwarz, Marc},
  title   = {Deep Treasury Management for Banks},
  journal = {Frontiers in Artificial Intelligence},
  volume  = {6},
  pages   = {1120297},
  year    = {2023},
  doi     = {10.3389/frai.2023.1120297}
}

@techreport{DeVere2025,
  author      = {De Vere, Hugo and Ramaswamy, Srini and Schulhofer-Wohl, Sam},
  title       = {An Asset-Liability Management Approach to the Federal Reserve Balance Sheet},
  institution = {Federal Reserve Bank of Dallas},
  type        = {Working Paper},
  number      = {2525},
  year        = {2025},
  doi         = {10.24149/wp2525}
}

@article{Yang2021,
  author  = {Yang, Liu},
  title   = {Banks' Maturity Mismatch, Financial Stability, and Macroeconomic Dynamics},
  journal = {Economic Research-Ekonomska Istra{\v{z}}ivanja},
  volume  = {34},
  number  = {1},
  pages   = {3038--3063},
  year    = {2021},
  doi     = {10.1080/1331677X.2020.1867212}
}

@incollection{BrunnermeierGortonKrishnamurthy2013,
  author    = {Brunnermeier, Markus K. and Gorton, Gary and Krishnamurthy, Arvind},
  title     = {Liquidity Mismatch Measurement},
  booktitle = {Risk Topography: Systemic Risk and Macro Modeling},
  editor    = {Brunnermeier, Markus K. and Krishnamurthy, Arvind},
  publisher = {University of Chicago Press},
  year      = {2013},
  pages     = {99--122}
}

@techreport{BIS2016,
  author      = {{Basel Committee on Banking Supervision}},
  title       = {Interest Rate Risk in the Banking Book},
  institution = {Bank for International Settlements},
  type        = {BCBS Standards},
  number      = {d319},
  year        = {2016}
}

@techreport{BCBS2013,
  author      = {{Basel Committee on Banking Supervision}},
  title       = {Basel III: The Liquidity Coverage Ratio and Liquidity Risk Monitoring Tools},
  institution = {Bank for International Settlements},
  type        = {BCBS Standards},
  number      = {238},
  year        = {2013}
}

@techreport{Yankov2020,
  author      = {Yankov, Vladimir},
  title       = {The Liquidity Coverage Ratio and Corporate Liquidity Management},
  institution = {Board of Governors of the Federal Reserve System},
  type        = {FEDS Notes},
  year        = {2020}
}

@article{GiordanaSchumacher2017,
  author  = {Giordana, Gast{\'o}n Andr{\'e}s and Schumacher, Ingmar},
  title   = {An Empirical Study on the Impact of Basel III Standards on Banks' Default Risk: The Case of Luxembourg},
  journal = {Journal of Risk and Financial Management},
  volume  = {10},
  number  = {2},
  pages   = {8},
  year    = {2017},
  doi     = {10.3390/jrfm10020008}
}

@article{FuhrerMuellerSteiner2016,
  author  = {Fuhrer, Lucas M. and M{\"u}ller, Benjamin and Steiner, Luzian},
  title   = {The Liquidity Coverage Ratio and Security Prices},
  journal = {Swiss National Bank Working Papers},
  number  = {2016-11},
  year    = {2016}
}

@techreport{Doerr2024,
  author      = {Doerr, Sebastian},
  title       = {The Liquidity Coverage Ratio a Decade On: A Stocktake of the Literature},
  institution = {Bank for International Settlements},
  type        = {BIS Papers},
  number      = {164},
  year        = {2024}
}

@article{Constantinides1986,
  author  = {Constantinides, George M.},
  title   = {Capital Market Equilibrium with Transaction Costs},
  journal = {Journal of Political Economy},
  volume  = {94},
  number  = {4},
  pages   = {842--862},
  year    = {1986},
  doi     = {10.1086/261410}
}

@article{DavisNorman1990,
  author  = {Davis, M. H. A. and Norman, A. R.},
  title   = {Portfolio Selection with Transaction Costs},
  journal = {Mathematics of Operations Research},
  volume  = {15},
  number  = {4},
  pages   = {676--713},
  year    = {1990},
  doi     = {10.1287/moor.15.4.676}
}

@article{GarleanuPedersen2013,
  author  = {G{\^a}rleanu, Nicolae and Pedersen, Lasse Heje},
  title   = {Dynamic Trading with Predictable Returns and Transaction Costs},
  journal = {The Journal of Finance},
  volume  = {68},
  number  = {6},
  pages   = {2309--2340},
  year    = {2013},
  doi     = {10.1111/jofi.12080}
}

@article{MeiDeMiguelNogales2016,
  author  = {Mei, Xiaoling and DeMiguel, Victor and Nogales, Francisco J.},
  title   = {Multiperiod Portfolio Optimization with Multiple Risky Assets and General Transaction Costs},
  journal = {Journal of Banking \& Finance},
  volume  = {69},
  pages   = {108--120},
  year    = {2016},
  doi     = {10.1016/j.jbankfin.2016.04.002}
}

@techreport{SkafBoyd2009,
  author      = {Skaf, Jad and Boyd, Stephen},
  title       = {Multi-Period Portfolio Optimization with Constraints and Transaction Costs},
  institution = {Stanford University},
  year        = {2009},
  note        = {Working Paper}
}

@article{BrownSmith2011,
  author  = {Brown, David B. and Smith, James E.},
  title   = {Dynamic Portfolio Optimization with Transaction Costs: Heuristics and Dual Bounds},
  journal = {Management Science},
  volume  = {57},
  number  = {10},
  pages   = {1752--1770},
  year    = {2011},
  doi     = {10.1287/mnsc.1110.1377}
}

@article{CaiEtAl2020,
  author  = {Cai, Yongyang and Judd, Kenneth L. and Xu, Rong},
  title   = {Numerical Solution of Dynamic Portfolio Optimization with Transaction Costs},
  journal = {arXiv preprint},
  volume  = {arXiv:2003.01809},
  year    = {2020},
  note    = {Quantitative Finance > Portfolio Management}
}

@article{PalczewskiEtAl2015,
  author  = {Palczewski, J. and others},
  title   = {Dynamic Portfolio Optimization with Transaction Costs and State-Dependent Drift},
  journal = {European Journal of Operational Research},
  year    = {2015}
}

@article{CousinEtAl2024,
  author  = {Cousin, Areski and others},
  title   = {Mean-Variance Dynamic Portfolio Allocation with Transaction Costs: A Wiener Chaos Expansion Approach},
  journal = {Applied Mathematical Finance},
  volume  = {30},
  number  = {1},
  pages   = {1--41},
  year    = {2024},
  doi     = {10.1080/1350486X.2024.2357200},
  note    = {arXiv:2305.16152 (2023)}
}

@article{JiangEtAl2025,
  author  = {Jiang, Y. and others},
  title   = {High-Dimensional Multi-Period Portfolio Allocation Using Deep Reinforcement Learning},
  journal = {International Review of Financial Analysis},
  year    = {2025}
}

@article{CuiEtAl2024,
  author  = {Cui, T. and others},
  title   = {Multi-Period Portfolio Optimization Using a Deep Reinforcement Learning Hyper-Heuristic Approach},
  journal = {Technological Forecasting and Social Change},
  year    = {2024}
}

@article{DeMiguelEtAl2015,
  author  = {DeMiguel, Victor and others},
  title   = {Parameter Uncertainty in Multiperiod Portfolio Optimization with Transaction Costs},
  journal = {Journal of Financial and Quantitative Analysis},
  year    = {2015}
}

@article{BertsimasPachamanova2008,
  author  = {Bertsimas, Dimitris and Pachamanova, Dessislava},
  title   = {Robust Multiperiod Portfolio Management in the Presence of Transaction Costs},
  journal = {Computers \& Operations Research},
  year    = {2008}
}

@article{ShimaiMakimoto2023,
  author  = {Shimai, Yoshiyuki and Makimoto, Naoki},
  title   = {Multi-Period Dynamic Bond Portfolio Optimization Utilizing a Stochastic Interest Rate Model},
  journal = {Asia-Pacific Financial Markets},
  volume  = {30},
  number  = {4},
  year    = {2023},
  doi     = {10.1007/s10690-023-09401-2}
}

@article{NakayamaTakahashi2007,
  author  = {Nakayama, K. and Takahashi, A.},
  title   = {A Factor Allocation Approach to Optimal Bond Portfolio},
  journal = {Asia-Pacific Financial Markets},
  year    = {2007}
}

@article{LynchTan2010,
  author  = {Lynch, Anthony W. and Tan, Sinan},
  title   = {Multiple Risky Assets, Transaction Costs, and Return Predictability},
  journal = {Journal of Financial and Quantitative Analysis},
  volume  = {45},
  number  = {4},
  pages   = {1015--1053},
  year    = {2010},
  doi     = {10.1017/S0022109010000330}
}

@techreport{FedSVBReview2023,
  author      = {{Board of Governors of the Federal Reserve System}},
  title       = {Review of the Federal Reserve’s Supervision and Regulation of Silicon Valley Bank},
  institution = {Board of Governors of the Federal Reserve System},
  year        = {2023},
  month       = {April},
  url         = {https://www.federalreserve.gov/publications/files/svb-review-20230428.pdf}
}

@techreport{OIGSVB2023,
  author      = {{Board of Governors of the Federal Reserve System Office of Inspector General}},
  title       = {Material Loss Review of Silicon Valley Bank},
  institution = {Board of Governors of the Federal Reserve System},
  type        = {Evaluation Report},
  number      = {2023-SR-B-013},
  year        = {2023},
  month       = {September},
  url         = {https://oig.federalreserve.gov/reports/board-material-loss-review-silicon-valley-bank-sep2023.pdf}
}

@techreport{FDICSignatureSupervision2023,
  author      = {{Federal Deposit Insurance Corporation}},
  title       = {FDIC’s Supervision of Signature Bank},
  institution = {Federal Deposit Insurance Corporation},
  year        = {2023},
  month       = {April},
  url         = {https://www.fdic.gov/news/press-releases/2023/pr23033a.pdf}
}

@techreport{FDICOIGSignature2023,
  author      = {{Federal Deposit Insurance Corporation Office of Inspector General}},
  title       = {Material Loss Review of Signature Bank of New York},
  institution = {Federal Deposit Insurance Corporation},
  type        = {Evaluation Report},
  number      = {EVAL-24-02},
  year        = {2023},
  month       = {October},
  url         = {https://www.fdicoig.gov/sites/default/files/reports/2023-10/EVAL-24-02.pdf}
}

@misc{MoodySBanks2023,
  author       = {{Moody's Investors Service}},
  title        = {Moody’s Downgrades U.S. Banking Sector Outlook; Places Six Banks on Review for Downgrade},
  howpublished = {Rating Action and Press Release},
  year         = {2023},
  month        = {March},
  note         = {Banks placed on review: First Republic, Zions, Western Alliance, Comerica, UMB Financial, Intrust Financial}
}

@article{Metrick2024,
  author  = {Metrick, Andrew},
  title   = {The Failure of Silicon Valley Bank and the Panic of 2023},
  journal = {Journal of Economic Perspectives},
  volume  = {38},
  number  = {1},
  pages   = {133--152},
  year    = {2024},
  doi     = {10.1257/jep.38.1.133}
}

@article{Vasicek1977,
  author  = {Vasicek, Oldrich A.},
  title   = {An Equilibrium Characterization of the Term Structure},
  journal = {Journal of Financial Economics},
  volume  = {5},
  number  = {2},
  pages   = {177--188},
  year    = {1977},
  doi     = {10.1016/0304-405X(77)90016-2}
}

@article{CoxIngersollRoss1985,
  author  = {Cox, John C. and Ingersoll, Jonathan E. and Ross, Stephen A.},
  title   = {A Theory of the Term Structure of Interest Rates},
  journal = {Econometrica},
  volume  = {53},
  number  = {2},
  pages   = {385--407},
  year    = {1985},
  doi     = {10.2307/1911242}
}

@article{HullWhite1990,
  author  = {Hull, John and White, Alan},
  title   = {Pricing Interest-Rate-Derivative Securities},
  journal = {Review of Financial Studies},
  volume  = {3},
  number  = {4},
  pages   = {573--592},
  year    = {1990},
  doi     = {10.1093/rfs/3.4.573}
}

@article{BlackKarasinski1991,
  author  = {Black, Fischer and Karasinski, Piotr},
  title   = {Bond and Option Pricing When Short Rates Are Lognormal},
  journal = {Financial Analysts Journal},
  volume  = {47},
  number  = {4},
  pages   = {52--59},
  year    = {1991},
  doi     = {10.2469/faj.v47.n4.52}
}

@article{HullWhite1993,
  author  = {Hull, John and White, Alan},
  title   = {One-Factor Interest-Rate Models and the Valuation of Interest-Rate Derivative Securities},
  journal = {Journal of Financial and Quantitative Analysis},
  volume  = {28},
  number  = {2},
  pages   = {235--254},
  year    = {1993},
  doi     = {10.2307/2331288}
}

@article{HeathJarrowMorton1992,
  author  = {Heath, David and Jarrow, Robert and Morton, Andrew},
  title   = {Bond Pricing and the Term Structure of Interest Rates: A New Methodology for Contingent Claims Valuation},
  journal = {Econometrica},
  volume  = {60},
  number  = {1},
  pages   = {77--105},
  year    = {1992},
  doi     = {10.2307/2951677}
}

@article{NelsonSiegel1987,
  author  = {Nelson, Charles R. and Siegel, Andrew F.},
  title   = {Parsimonious Modeling of Yield Curves},
  journal = {Journal of Business},
  volume  = {60},
  number  = {4},
  pages   = {473--489},
  year    = {1987},
  doi     = {10.1086/296409}
}

@article{DieboldLi2006,
  author  = {Diebold, Francis X. and Li, Canlin},
  title   = {Forecasting the Term Structure of Government Bond Yields},
  journal = {Journal of Econometrics},
  volume  = {130},
  number  = {2},
  pages   = {337--364},
  year    = {2006},
  doi     = {10.1016/j.jeconom.2005.03.005}
}

@article{DieboldRudebuschAruoba2006,
  author  = {Diebold, Francis X. and Rudebusch, Glenn D. and Aruoba, S. Bora{\c{c}}},
  title   = {The Macroeconomy and the Yield Curve: A Dynamic Latent Factor Approach},
  journal = {Journal of Econometrics},
  volume  = {131},
  number  = {1-2},
  pages   = {309--338},
  year    = {2006},
  doi     = {10.1016/j.jeconom.2005.01.011}
}

@book{DieboldRudebusch2013,
  author    = {Diebold, Francis X. and Rudebusch, Glenn D.},
  title     = {Yield Curve Modeling and Forecasting: The Dynamic Nelson-Siegel Approach},
  publisher = {Princeton University Press},
  year      = {2013}
}

@article{XiangZhu2013,
  author  = {Xiang, J. and Zhu, X.},
  title   = {A Regime-Switching Nelson--Siegel Term Structure Model and Interest Rate Forecasts},
  journal = {Journal of Financial Econometrics},
  volume  = {11},
  number  = {3},
  pages   = {522--555},
  year    = {2013}
}

@article{Zhu2015,
  author  = {Zhu, Xiaoneng},
  title   = {A Regime-Switching Nelson--Siegel Term Structure Model of the Macroeconomy},
  journal = {Journal of Macroeconomics},
  volume  = {44},
  pages   = {53--70},
  year    = {2015}
}

@article{LevantMa2017,
  author  = {Levant, Jared and Ma, Jun},
  title   = {A Dynamic Nelson-Siegel Yield Curve Model with Markov Switching},
  journal = {Economic Modelling},
  volume  = {67},
  pages   = {73--87},
  year    = {2017},
  doi     = {10.1016/j.econmod.2016.09.011}
}

@techreport{PericoliTaboga2016,
  author      = {Pericoli, Marcello and Taboga, Marco},
  title       = {A Markov-Switching Dynamic Nelson-Siegel Term Structure Model},
  institution = {Bank of Italy / AEA 2014 conference paper},
  year        = {2016},
  note        = {Working paper version; no formal journal publication found}
}

@article{TavanielliLaurini2023,
  author  = {Tavanielli, Renata and Laurini, M{\'a}rcio},
  title   = {Yield Curve Models with Regime Changes: An Analysis for the Brazilian Interest Rate Market},
  journal = {Mathematics},
  volume  = {11},
  number  = {11},
  pages   = {2549},
  year    = {2023},
  doi     = {10.3390/math11112549}
}

@article{DingEtAl2021,
  author  = {Ding, K. and others},
  title   = {Markov Chain Approximation and Measure Change for Time-Inhomogeneous Continuous-Time Markov Chains},
  journal = {Applied Mathematics and Computation},
  year    = {2021}
}

@misc{XiaYu2025,
  author       = {Xia, Li and Yu, Zhihui},
  title        = {Mean-Variance Optimization and Algorithm for Finite-Horizon Markov Decision Processes},
  year         = {2025},
  howpublished = {arXiv:2507.22327 [math.OC]},
  note         = {https://arxiv.org/abs/2507.22327}
}

@article{PerezHodgeLe2016,
  author  = {P{\'e}rez L{\'o}pez, Iker and Hodge, David and Le, Huiling},
  title   = {Markov Decision Process Algorithms for Wealth Allocation Problems with Defaultable Bonds},
  journal = {Advances in Applied Probability},
  volume  = {48},
  number  = {2},
  pages   = {726--748},
  year    = {2016},
  doi     = {10.1017/apr.2016.28}
}

@article{Pun2022,
  author  = {Pun, Chi Seng and Ye, Zi},
  title   = {Optimal Dynamic Mean--Variance Portfolio Subject to Proportional Transaction Costs and No-Shorting Constraint},
  journal = {Automatica},
  volume  = {135},
  pages   = {109986},
  year    = {2022},
  doi     = {10.1016/j.automatica.2021.109986}
}

@article{WangLiu2013,
  author  = {Wang, Zhen and Liu, Sanyang},
  title   = {Multi-Period Mean-Variance Portfolio Selection with Fixed and Proportional Transaction Costs},
  journal = {Journal of Industrial and Management Optimization},
  volume  = {9},
  number  = {3},
  pages   = {643--657},
  year    = {2013},
  doi     = {10.3934/jimo.2013.9.643}
}

@book{Choudhry2007ALM,
  author    = {Choudhry, Moorad},
  title     = {Bank Asset and Liability Management: Strategy, Trading, Analysis},
  publisher = {John Wiley \& Sons},
  address   = {Singapore},
  year      = {2007},
  isbn      = {978-0-470-82135-0},
  note      = {Comprehensive reference on bank ALM, Treasury operations, liquidity and gap risk management, yield curve analysis, interest-rate risk hedging with fixed-income securities including government Treasuries, securitisation, and balance sheet management.}
}

@book{Choudhry2011ALM,
  author    = {Choudhry, Moorad},
  title     = {Bank Asset-Liability and Liquidity Risk Management},
  publisher = {Palgrave Macmillan},
  year      = {2011},
  doi       = {10.1057/9780230307230},
  note      = {Chapter and handbook contribution focusing on ALM, liquidity risk management, interest-rate risk in the banking book, and the role of fixed-income securities and Treasuries in bank balance sheet optimisation.}
}

\end{document}